\documentclass[11pt]{amsart}
\usepackage{amsaddr}
\pdfoutput=1
\usepackage{graphicx}
\usepackage{chemformula} 
\usepackage[T1]{fontenc} 
\usepackage[bitstream-charter]{mathdesign}

\usepackage{hyperref}
\usepackage{amsfonts,amsmath,amssymb}
\usepackage{microtype}
\usepackage[version=3]{mhchem}
\usepackage{nicefrac}
\usepackage{booktabs}

\title[Vibrationally Assisted Heterogeneous Electron Transfer]
  {A femtosecond time resolved view of vibrationally assisted electron transfer across the metal/aqueous interface}
\author{Zhipeng Huang$^{\dag\star}$, Manuel Bridger$^{\dag\star}$, Oscar A.\ Naranjo-Montoya$^{\dag}$, Alexander Tarasevitch$^{\dag}$, Uwe Bovensiepen$^{\dag}$, Yujin Tong$^{\dag}$, R.\ Kramer Campen$^{\dag}$}
\email{uwe.bovensiepen@uni-due.de}
\email{yujin.tong@uni-due.de}
\email{richard.campen@uni-due.de}
\address{$^{\dag}$Faculty of Physics and Center for Nanointegration Duisburg-Essen (CENIDE), University of Duisburg-Essen, 47048, Duisburg, Germany}
\address{$^{\star}$ \emph{equal contribution}}

\begin{document}
\maketitle

\begin{abstract}
Understanding heterogeneous charge transfer is crucial if we are to build the best electrolyzers, fuel cells and photoelectrochemical water splitting devices that chemistry allows. 
Because the elementary processes involved have timescales ranging from femto- to milliseconds, direct simulation is not generally possible.
Model Hamiltonian approaches thus have a crucial role in gaining mechanistic insight. 
Current generations of such theories describe a reactant(s) or product(s) that interacts with electrolyte via a single effective interaction. 
Such approaches thus obscure the extent to which \emph{particular} solvent fluctuations influence charge transfer. 
Here we demonstrate experimentally that for a prototypical system, a ferrocene terminated alkane thiol self-assembled monolayer (SAM) on gold in contact with aqueous electrolyte, charge transfer from the Au to the ferrocene can be induced by vibrational excitation of the ferrocene aromatic CH.
Intriguingly the energy of the aromatic CH vibration, 0.38 eV, is a large fraction of the effective solvent interaction strength inferred for the ferrocene/ferrocenium system in prior electrochemical studies: 0.85 eV. 
Our results thus demonstrate the coupling of charge transfer to a \emph{specific} solvent motion and more generally imply that solvent may 
affect reduction/oxidation rates in electrocatalysis by coupling to a few distinct solvent motions.
Identifying these motions is crucial in rationalizing trends in reactivity with change in electrolyte and thus in pursuing \emph{electrolyte engineering} from first principles. 
\end{abstract}

\section{Introduction}
Electron transfer across the (solid metallic) electrode/ (liquid) electrolyte interface underlies much of the chemistry necessary for a sustainable global energy economy \cite{zhu21b,gle22}.
Thus building the best possible energy conversion or storage devices requires understanding transfer mechanism. 
However, while well established observations clearly illustrate the dependence of transfer rate on solid phase, electrolyte and surface structure for a variety of important reactions, inferring molecular-scale mechanisms self-consistent with such trends has proven challenging \cite{sch15c,nar22}.
 
In principle suitably referenced ab-initio molecular dynamics simulation offers insight into such questions \cite{dei21}. 
In practice, however, required system sizes (hundreds to thousands of solute molecules) and characteristic timescales of interest (fs to millsecond) are too large to permit direct \emph{numerical experiments} for all but the simplest systems \cite{abi20,gro23}. 
An appropriate model Hamiltonian theory is thus important both in extracting the essential physics for systems amenable to direct simulation and rationalizing experimental trends in activity for the great majority of systems that are not \cite{san22}.
The necessary elements of such a theory depends on adsorbate/metal interaction strength \cite{sch10,san12,san12b}.

At very weak interaction strength electron transfer is non-adiabatic.
In this regime transfer rates are too small to be useful electrochemically and prediction of rates is possible using a perturbation treatment \cite{dog72}. 
With increasing interaction strength one enters a `weakly adiabatic' regime in which electron transfer rates are insensitive to adsorbate/metal interaction strength and individual transfer events are fast relative to solvent configurational fluctuations. 
In this regime semi-classical Marcus-Hush theory \cite{mar65,hus61}, its quantum mechanical generalization \cite{dog72} and its extension to electrocatalysis \cite{san22,sch86} apply. 
For strongly interacting species adsorption results in formation of hybridized electronic states and the notion of \textit{electron transfer} may not, as such, apply. 
Essentially all systems of interest in electrocatalysis or energy conversion have adsorbate/metal interaction strength between the \textit{weakly adiabatic} and \emph{strongly interacting} regimes.
Troublingly, as laid out by Schmickler and Santos \cite{sch10}, there is no \emph{general} theory that describes electron transfer in these situations: \textit{e.g.}\ current calculated rates for the well studied hydrogen evolution on Pt(111) may differ by 10$^{4}$ from those found in experiment \cite{he18}. 
Here we begin by describing Marcus-Hush theory with a particular focus on its conceptually well defined, but difficult to parse, description of the solvent, and then demonstrate results from a novel experimental approach that allows specific insight into the role of solvent in electron transfer for a sample in the \textit{catalytic regime}. 

In Marcus-Hush theory electron transfer occurs adiabatically when solvent configuration permits \cite{sch10}. 
If the reactant species is assumed to interact with its solvent under a harmonic potential and one integrates over the solid's density of states\footnote{Assumed here to be metallic and thus have a continuum of energy levels $\epsilon$(eV)}, the rate constants for reduction(/oxidation) are (in which the upper sign indicates reduction, the lower oxidation) \cite{lab12}:
	\begin{equation}
	k_{\text{red/ox}} = \int^{\infty}_{-\infty}\text{A}_{\text{red/ox}}(\epsilon)\frac{\exp\left(\nicefrac{-\Delta G^{\dag}_{\text{red/ox}}(\epsilon)}{kT}\right)}{1 + \exp\left(\nicefrac{\mp e(\epsilon - \phi)}{kt}  \right)} d\epsilon
	\end{equation}
where $e$ is the charge of the electron, $k$ Boltzmann's constant, $\text{A}_{\text{red/ox}}(\epsilon)$  the pre-exponential factor (which is a function of coupling between the metal and electroactive states as well as the metal density of states), and $\Delta G^{\dag}_{\text{red/ox}}$ is the free energy of the transition state.
Given a parabolic potential energy surface in both the reductant and oxidant, the free energies of the two transition states are:
	\begin{equation}\label{e:transition_states}
		\Delta G^{\dag}_{\text{red/ox}} (x)  =  \frac{\lambda^{*}}{4} \left(1 \pm \frac{\eta + x}{\lambda^{*}} \right)^{2} 
	\end{equation}
in which $\lambda^{*}$ is the dimensionless reorganization energy, \textit{i.e.} $\lambda^{*} = \frac{\text{F}\lambda}{\text{RT}}$, $\eta$ the dimensionless overpotential and $x$ a dimensionless factor accounting for the solid density of states.
In such a description electron transfer rates between a solid and an adsorbed molecule are a function of bulk solid electronic structure and a single, thermally averaged, solvent reorganization energy ($\lambda$).
From a microscopic perspective the solvent is composed of $\nu$ harmonic oscillators, each with frequency $\omega_{\nu}$, linearly coupled, with coupling constant $g_{\nu}$, to the electron transfer coordinate \cite{kop13}. 
Thus $\lambda$ can be expressed,
\begin{equation}\label{e:lambda}
	\lambda = \frac{1}{2}\sum_{\nu}\hbar \omega_{\nu}g_{\nu}^{2}
\end{equation} 
Identifying each of these solvent oscillators and the strength of their coupling to the charge transfer coordinate has not been possible. 
Because only $\lambda$ enters calculation of the rate, the solvent influence on reaction rate is often taken to be a single effective harmonic oscillator.

Historically much of the effort in electrocatalysis, whether experimentally\cite{smo03,mcg15,gre23} or theoretically\cite{san22,gre06,li19d}, has been directed towards finding new solid catalyst materials.  
More recently it has become clear that the selectivity and activity of particular catalysts are often dramatic functions of the adjoining electrolyte \cite{hua21}.
Clearly such observations cannot be rationalized by solvent acting on the charge transfer coordinate as a single \textit{effective} harmonic oscillator: insight into the right hand side of \autoref{e:lambda} is required.
 
Put physically, we wish to understand the manner in which molecular motion influences charge transfer from a bulk solid to an adjoining molecular phase. 
This question has been experimentally addressed for inorganic/organic heterostructures by performing experiments in which above band gap excitation of the inorganic phase induces a photocurrent that is modulated by excitation of particular vibrations in the organic \cite{wri12,bak16}. 
Extending this approach to the solid/liquid interface requires initiating electron transfer at a well defined time and measuring the current or voltage associated with this transfer in the presence and absence of a vibrational excitation. 
Here we overcome this challenge using a two photon photovoltage (2PPV) approach in which a femtosecond optical pump pulse quasi-instantaneously creates a subpopluation of electrons with increased energy (analogous to increasing electron chemical potential) in a metal electrode and the photovoltage (under open circuit conditions) is measured as a function of femtosecond time-resolved delay to a second pulse of variable photon energy. 
By changing the photon energy of the second pulse in (and out of) resonance with a particular solvent vibration we characterize the influence of a  specific solvent motion on charge transfer. 
In this study we apply this approach to Ferrocene terminated alkane thiol self-assembled monolayers on Au. 

Such systems are characterized by their extreme stability, reversible one electron transfer redox chemistry and facilitation of rapid interfacial charge transfer \cite{uos91,kon95}. 
Binary self-assembled monolayers (SAMS) containing a mixture of Ferrocene terminated and nonterminated alkane thiols are a weakly-adiabatic system well described by Marcus-Hush theory \cite{chi91,rud13}.
Hirata and coworkers have shown, using ultraviolet and two photon photoelectron spectroscopy, that monocomponent Ferrocene terminated alkane-thiol SAMS result in a defect ridden monolayer in which electronic coupling between the Ferrocene and Au is increased relative to the binary\cite{hir13}.
Such increased coupling is consistent with a scenario in which electron transfer in these systems is of the \emph{catalytic regime} type .
In this study we explore the relationship of vibrational excitation to charge transfer in a such a defect-rich Ferrocene terminated alkane thiol SAM.
By measuring the 2PPV response as a function of photon energy and delay between two optical pulses we show that charge transfer from Au to ferrocene is enhanced in the presence of mid-infrared pulses that excite the aromatic C-H (but not of the aliphatic C-H) stretch vibrations. 
These measurements allow us to experimentally demonstrate the presence of vibration assisted charge transfer across the Au/Ferrocene SAM/aqueous solution interface for a monocomponent Ferrocene terminated monolayer.  
To our knowledge the results presented here are the first experimental observations of the coupling of a solvent vibration to heterogeneous electron transfer: the first demonstration that it is even possible to understand the right hand side of \autoref{e:lambda}. 
This experimental capability is crucial in understanding exactly how solvent controls electron transfer in the \emph{weakly adiabatic} and \textit{catalytic regimes}. 
Because of the fundamental nature of the effect we explore, and the general applicability of the mixed optoelectronic method we employ, these results should be of general interest.

\section{Methods}
\label{sec:Experimental method}
\subsection{Laser System for 2PPV Measurements}
\begin{figure}[h]
	\begin{center}
		\includegraphics[width=\textwidth]{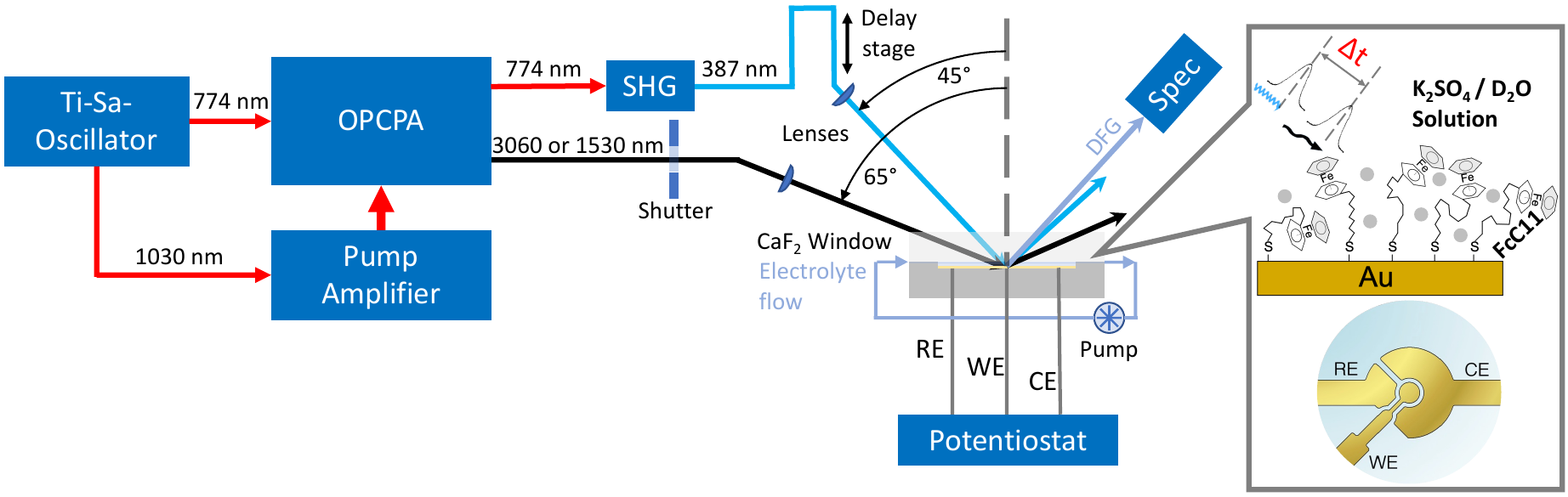}
		\caption{Schematic layout of the laser system, spectroelectrochemical cell and a microscopic depiction. SHG: Second harmonic generation; RE, CE, WE: Reference, Counter and Working Electrodes; Spec: Spectrometer (SpectraPro HRS-300).\label{fig:setup}}
	\end{center}
\end{figure}

To perform the 2PPV experiments, we employed the output from several stages of a home-built four stage optical parameteric chirped pulse amplifier (OPCPA) laser system described previously  
and shown schematically in \autoref{fig:setup} \cite{Bridger:19}.
In brief, a Ti-Sapphire Oscillator running at 80 MHz acts as a seed for both the pump amplifier and the four stage OPCPA chain. 
Within the OPCPA the seed, at 774 nm, is amplified, pulses centered at 1530 and 3060 nm produced and the repetition rate decreased to 100 Hz.
The 387 nm pulse train was generated by frequency doubling the 774 nm output in a BBO crystal as indicated. 
To perform an experiment the 774 or 387 nm pulse train was delayed with respect to the 3060 (or 1530) nm.
Pulse lengths and powers employed for all photon energies, and spectra and spatial profiles of the 387 and 3060 nm, are shown in the Supporting Information.

\subsection{FcC$_{11}$ SAM Preparation and Sample Cell}
95 \% pure FcC$_{11}$, \textit{i.e.}\ 11-(Ferr\-ocenyl)-undecanthiol, was purchased from Sigma-Aldrich and used as received. 
Gold working, counter and pseudo-reference electrodes (5 nm Cr/200 nm Au) were prepared by electron beam heated physical vapor deposition on top of a silica plate (see inset of \autoref{fig:setup}).
After deposition the electrodes were sonicated in acetone, methanol and ultrapure water (Milliure, 18.2 M$\Omega$) for 10 minutes each, cleaned in UV generated-Ozone (BioForce Nanosciences) for 20 minutes, rinsed with methanol and dried in Ar. 
The cleaned electrodes were then immersed in a 2 mM FcC$_{11}$ methanol solution (Acros Organics, 99.9 \% purity) for 2 days. 
Before assembling the electrodes into the (spectro)electrochemical cell, they were rinsed with ultrapure water and dried with N$_{2}$. 
We constructed a thin layer sample cell using the FcC$_{11}$ functionalized Au working electrode (see \autoref{fig:setup}), a \ce{CaF2} window, a 50 $\mu$m teflon spacer and a 0.5 mol/L \ce{K2SO4} (Alfa Aesar, >99.0 \% purity)/\ce{D2O} solution. 
Photovoltages were detected by a Biologic SP-200 potentiostat sampling at 25 Hz. 

We characterized the structure of the FcC$_{11}$ monolayer electrochemically, by conducting cyclic voltammetry using the potentiostat and three electrode cell shown in \autoref{fig:setup}, and optically using vibrationally resonant sum frequency generation (VSFG) spectroscopy. 
To conduct the VSFG measurement we employed a homebuilt spectrometer and separate laser system described in the Supporting Information.

\section{Results and Discussion} 
\subsection{Confirming the FcC$_{11}$ SAM is Stable and Disordered}
Self-assembled monolayers composed of mixtures of ferrocene terminated alkanethiols (FcC$_{n}$) and unsubstituted alkanethiols (C$_{n}$) have received wide attention for their potential application in sensors and molecular electronics and because of the ideal Marcus-Hush redox behavior of their one electron reduction (\textit{i.e.}\ Ferrocene to Ferrocenium) \cite{uos91,chi91,kon95}.
SAMS that contain mixtures of FcC$_{11}$ and C$_{11}$ on polycrystalline Au, in which the former component is dilute, \textit{i.e.}\ $< 0.6\times 10^{10}$ mol/cm$^{2}$, are stable in acidic electrolyte and show a single, reversible, redox feature at $\approx 0.2$ V v.\ SCE \cite{rud13}.
Increasing the coverage of the FcC$_{11}$ component (or changing anion \cite{won20}) leads to shifting and splitting of this current feature.
Prior two photon photoemission measurements have found that a monocomponent, relatively disordered SAM results in Ferrocene head groups that, on average, interact more strongly with Au than the ideal case (in which alkane chains are each in all trans conformations) suggesting that disordered FcC$_{11}$ SAMS are a model heterogeneous electron transfer system in the \textit{catalytic regime} \cite{hir13}.

Performing a 2PPV experiment requires a FcC$_{11}$ SAM that remains adsorbed, uncontaminated and has a stable energy of the Ferrocene/Ferrocenium redox couple over the course of the experiment. 
Our sample cell requires the use of a \ce{CaF2} IR transmissive window in contact with electrolyte.
\ce{CaF2} is soluble in acid but relatively insoluble at circumneutral pH. 
As shown in the Supporting Information, preparing a monocomponent FcC$_{11}$ monolayer in the manner described and then performing cyclic voltammetry employing a 0.5 mol/L circumneutral pH \ce{K2SO4}/\ce{D2O} solution as electrolyte results in a voltammogram that shows a single, reversible current feature stable over the 10-12 hours necessary to perform a 2PPV measurement. 
Integrating the resulting current features and estimating the electrochemically active surface area \cite{mcc13} suggests a surface density of  $2.04\times 10^{10}$ mol (of FcC$_{11}$)/cm$^{2}$. 
Prior measurements within a bicomponent SAM suggest that at such FcC$_{11}$ densities interaction between neighboring Ferrocene groups is relatively weak and Ferrocene closer to Au than in the bicomponent well packed system \cite{rud13}. 
If this scenario is correct the SAM should also be disordered: individual methylene groups in the alkane tails oriented both cis and trans with respect to their neighbors.

\begin{figure}[htbp]
	\begin{center}
		 \includegraphics[width=0.55\textwidth]{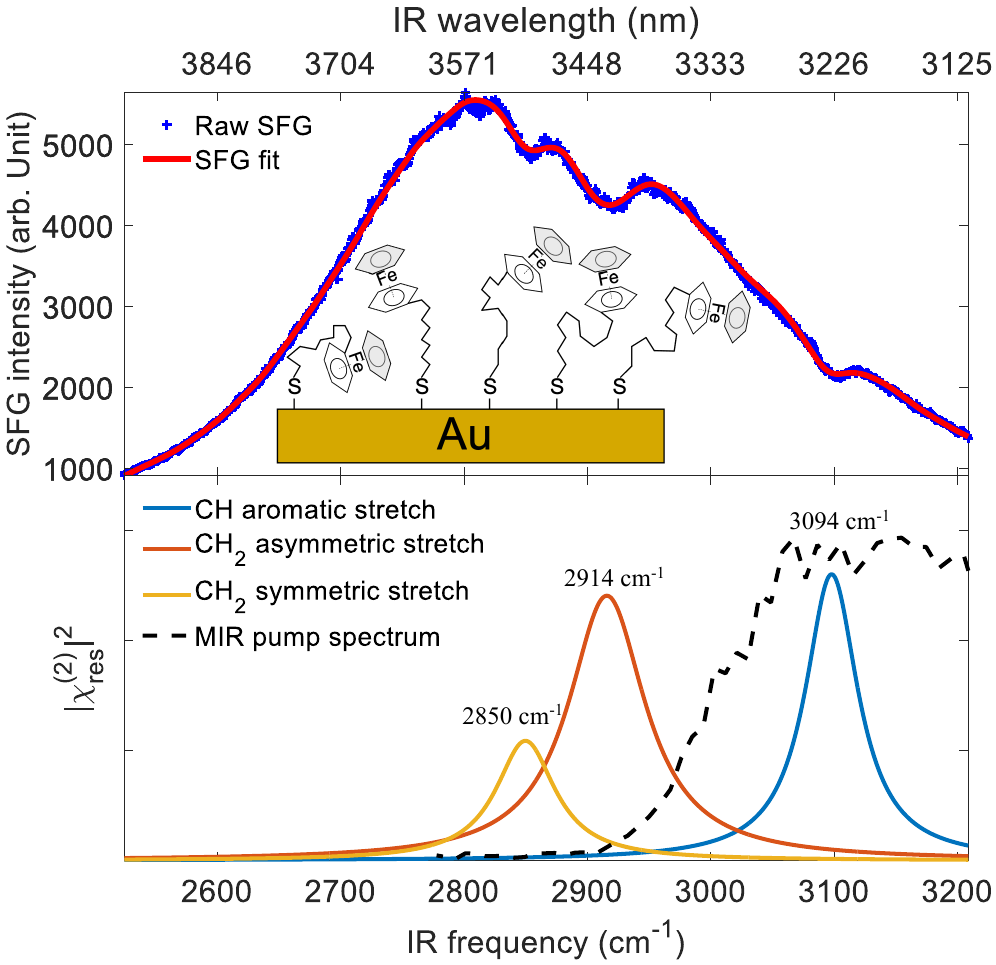}
		 \caption{Vibrationally resonant sum frequency spectrum of an FcC11 SAM on a gold surface in air. The large apparent peak dominating the signal is the non-resonant spectral response of the Au. Its frequency dependence is the result of the frequency dependent infrared intensity. CH resonant features interfere with this background and appear as small dips. \label{fig:SFG}}
	 \end{center}
\end{figure}

VSFG spectroscopy is prohibited in media with local inversion symmetry by its symmetry selection rules\cite{she89,lam05}. 
Much prior work on surfactants containing alkane tails and alkane thiol SAMS has demonstrated that well-packed surfactants and SAMS form relatively defect free interfacial \textit{structures} in which all pairs of neighboring methylene carbon atoms (\textit{i.e.}\ \ce{R2CH2} groups) have trans configurations \cite{rok03b,sun10}. 
Because such trans oriented \ce{CH2} groups have inversion symmetry with their neighbors the VSFG \ce{CH2} spectral response in these systems is a quantitative probe of structural disorder (number of cis oriented \ce{CH2} groups).
\autoref{fig:SFG}(top) shows a VSFG spectrum, plotted as a function of incident IR photon energy, of the FcC$_{11}$ monolayer prepared as described above in air.
The spectrum shows a broad background, the sum frequency response of the Au surface due to the frequency-dependent IR source, with several small dips corresponding to CH stretch vibrations. 
Fitting a line shape model to this data, see Supporting Information for details, enables the extraction of the spectral amplitudes shown in \autoref{fig:SFG}(bottom). 
Comparison with prior work allows assignment of the features at 2850 and 2914 cm$^{-1}$ to the \ce{CH2} symmetric and asymmetric stretches (originating from the disordered C$_{11}$ tails) and the feature at 3094 cm$^{-1}$ to the aromatic CH stretch vibration (originating from the Ferrocene head group) \cite{lu04}. 
Clearly these spectra are consistent with a disordered monolayer. 
As also illustrated in \autoref{fig:SFG}(bottom), the mid-IR pulse employed in the 2PPV measurements  largely overlaps only with the aromatic CH stretch. 

\subsection{2PPV results: Control Experiments and Understanding our Signal}
\begin{figure}[htbp]
\centering
\includegraphics[width=0.9\textwidth]{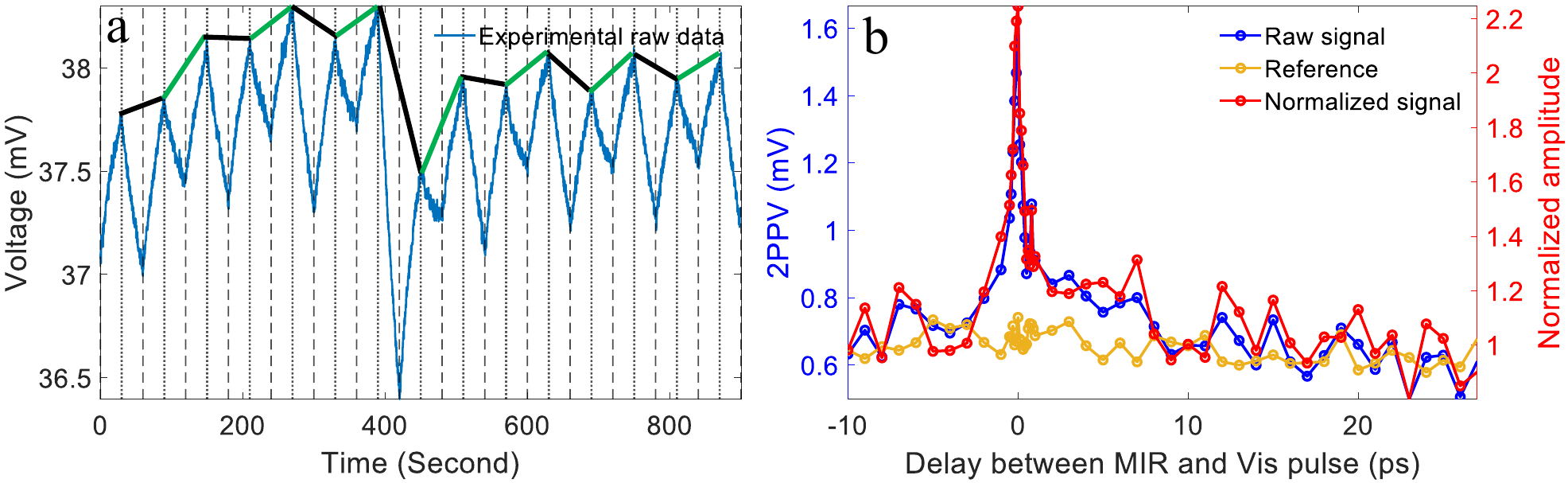}
\caption{(a) Raw data of the two photon (3060~nm and 774~nm) pump - photovoltage probe scheme read off the potentiostat. Each upper point in the blue saw tooth pattern occurs when the shutter in the 3060 nm beam is opened, each lower point when it is closed. Each `triangle' is the 2PPV response at a particular delay. Normalized signals, shown in panel (b), are calculated by taking the difference between a data point (the dots forming the blue lines in panel (a)) and the base of the `triangle' (black, data, or green, reference, lines). The points in (b) are the result of averaging all voltages in each triangle. Negative delay times indicate the 774 nm pulse arrives before the 3060 nm.}
\label{fig:774nm}
\end{figure}

\begin{figure}[!htbp]
\centering
\includegraphics[width=0.6\textwidth]{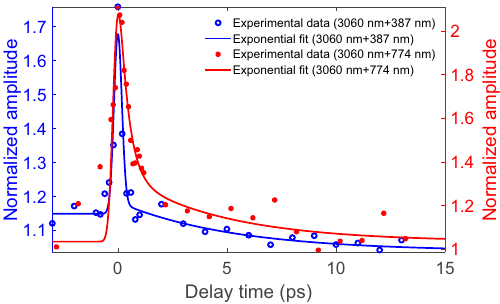}
\caption{Two photon pump (3060 nm + 387 nm shown in blue, 3060 nm + 774 nm shown in yellow) - photovoltage probe normalized signals. The results were fitted to a convolution function of a double exponential function with a Gaussian function, whose width came from the cross correlation measurement of two pump pulses.
Negative delay times indicate either the 384 or 774 nm pulse precedes the 3060 nm.} 
\label{fig:774nm_387nm}
\end{figure}

\begin{figure}[htp]
	\begin{center}
		\includegraphics[width=\textwidth]{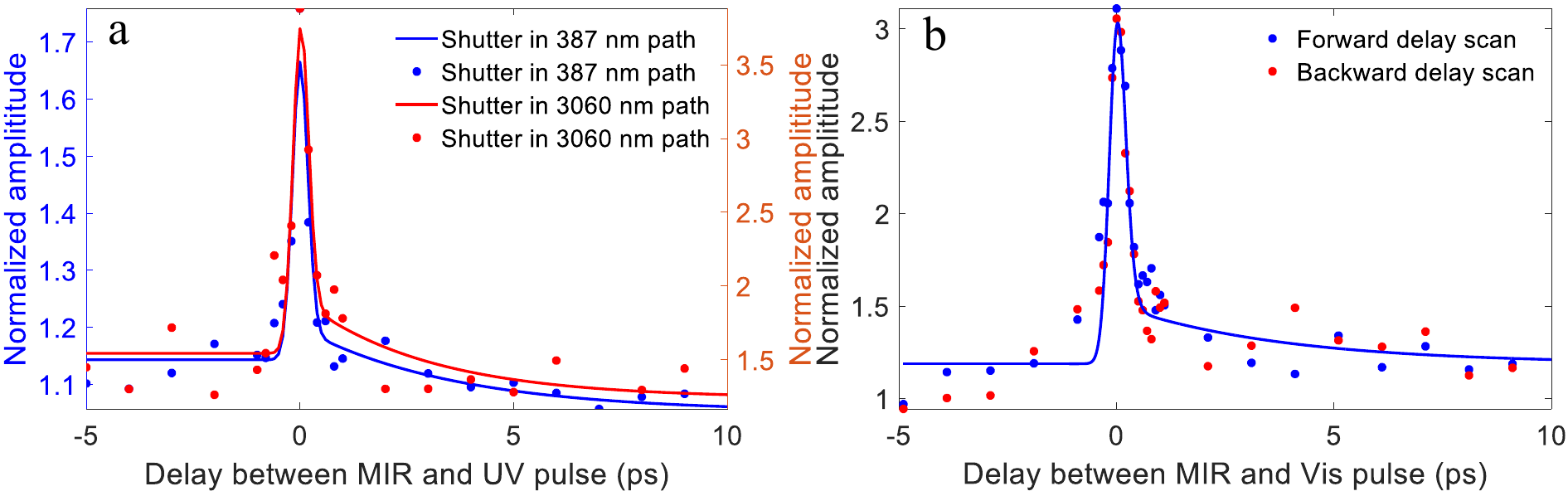}
		\caption{(a) The normalized 2PPV signal for an experiment combining pulse trains with wavelengths of 387 nm and 3060 nm. The blue points shows an experiment in which the shutter was placed in the 387 nm beam path (\textit{i.e.}\ the sample was continuously exposed to the 3060 nm pulse train) and the orange points the opposite (lines are guides to the eye). Clearly the placement of the shutter does not change the normalized signal. (b) A 774 nm + 3060 nm experiment in which delays were collected from negative to positive (\textit{i.e.}\ starting with 774 \emph{before} 3060) and positive to negative. Clearly the order in which delay points were sampled does not change the underlying dynamics.  \label{fig:2PPV_Control}}
	\end{center}
\end{figure}

Before performing the measurement a cyclic voltammogram (CV) is collected within the spectroelectrochemical cell, and the sample is allowed to stabilize at its open circuit voltage (OCV). 
As shown on the CV in the Supporting Information the OCV under these conditions is cathodic of the Ferrocene/Ferrocenium redox peak: the monolayer is dominated by Ferrocene before laser pulse arrival.
\autoref{fig:774nm}a shows the measured data for a 2PPV experiment with 3060 and 774 nm beams performed by placing a shutter in the 3060 nm beam, opening it, measuring the voltage, closing the shutter, changing the delay between the two pulses and repeating.
Each \emph{open} and \emph{closed} state is 60 seconds long (the delay is changed while the shutter is closed). 
A reference signal (in which the 774 pulse is 150 ps before the 3060 nm) was measured after each delay. 
It is clear from inspection of \autoref{fig:774nm}a that opening the shutter leads (as it does for all experiments in which an effect is observable) to a \emph{decrease} in OCV, \textit{i.e.}\ the 2PPV signal results from a redistribution of charge towards the solution. 
The 2PPV signal is extracted by integrating the area of the triangle formed for the shutter open and the following shutter closed period (the triangle composed of a two blue sides and a black or green bottom in \autoref{fig:774nm}a).
As we show in the Supporting Information, dividing the measured photovoltage response by the signal at -150 ps results in a normalized 2PPV response that is insensitive to the manner in which the potentiostat signal is integrated and the length of time the shutter is open/closed. 
\autoref{fig:774nm}b shows the normalized 2PPV signal for the experiment in \autoref{fig:774nm}a. 
Three qualitative features of the data are apparent from inspection: (1) there is no signal when the 774 nm pulse arrives at the electrode surface \emph{before} the 3060 nm (\textit{i.e.}\ negative delay times, see also \autoref{fig:774nm_387nm} and \autoref{fig:2PPV_Control}) (2) near time 0 a large 2PPV signal is apparent that has a lifetime of a few hundred femtoseconds and (3) a signal that persists for > 1 picosecond.
Quantifying the dynamics apparent in the signal via a double exponential fitting gives $\tau_{\text{fast}} = 405 \pm 165$ fs and $\tau_{\text{slow}} = 4.7 \pm 3$ ps (fitting results are shown in \autoref{fig:774nm_387nm}). 

As shown in \autoref{fig:774nm_387nm}, experiments that combine 3060 and 387 nm beams yield signals with quantitatively similar dynamics: $\tau_{\text{fast}} = 46 \pm 13$ fs and $\tau_{\text{slow}} = 5.3 \pm 4.2$ ps. 
Other pulse combinations, 774 nm + 1530 nm and 387 nm + 1530 nm, have no 2PPV signal (see Supporting Information).
In this experimental scheme our goal is to probe a system in which the creation of hot electrons in Au with the 387 nm or 774 nm beam leads to electron transfer to the FcC$_{11}$ head group the extent of which is enhanced in the presence of excitation of the aromatic CH stretch and relaxes between each laser pulse pair. 
As we show next, a number of control experiments strongly suggest we have achieved this goal. 

While the physical processes that control the two photon open circuit photovoltage are not well explored in the literature changes in temperature or solution phase composition are known to influence the OCV in other aqueous electrochemical contexts \cite{yan19d,Tong20,del21}. 
To confirm that accumulation of heat or solution phase products does not effect our 2PPV observations we performed two types of experiments: (i) an experiment in which the 2PPV signals with the shutter placed in the 387 nm and 3060 nm were compared (\autoref{fig:2PPV_Control}a) and (ii) sampling the delay points from positive to negative times and negative to positive (\autoref{fig:2PPV_Control}b). 
\autoref{fig:2PPV_Control} clearly illustrates that our 2PPV signal is insensitive to such changes.
Taken together both sets of results clearly suggest that our 2PPV signal is the result of a process that is reversible on timescales faster than the laser pulse repetition rate.
To assign this process requires further consideration of the photon energies of the set up and the system. 
We employ four photon energies in this experiment: 387 nm (3.20 eV), 774 nm (1.60 eV), 1530 nm (0.81 eV) and 3060 nm (0.41 eV). 
The only vibration in the system that significantly spectrally overlaps with any of the incident pulses is the aromatic CH stretch of the Ferrocene head group (with the 0.41 eV). 
In both crystalline form and in solution, Ferrocene has an absorption centered at $\approx$ 440 nm (2.82 eV) while the Ferrocenium cation absorbs at 610 nm (2.03 eV) with a broad shoulder towards shorter wavelengths \cite{arm67,soh70,scu09,rab13}.
These features can be altered significantly by adsorption on metals. 
The reflectance of Au is $\approx 40$ \% at wavelengths from 300-500 nm (a consequence of the 5d $\rightarrow$ 6s intraband transition), drastically increases to 95 \% between 500 and 700 nm and is wavelength independent beyond 700 nm \cite{dor65}. 
The relatively high density of states in Au, combined with the relatively small number of Ferrocene head groups within the optical focal volume(s), suggests that, at least the 387 nm (3.20 eV) pulses interact directly with the Au substrate. 
This supposition has been experimentally verified in 2 photon photoemission measurements of FcC$_{11}$ monolayers of varying order \cite{hir13}.
We thus conclude that the dynamics we observe are driven by the interaction of the 3.20 eV and 1.60 eV pulses with Au electrons (the former by a one photon the later by a two photon mechanism) and the 0.41 eV with the aromatic CH.

In this experiment, we detect a small modulation of a relatively large OCV. 
Given that our laser emits subpicosecond pulses at 100 Hz, the potentiostat records at 25 Hz and we did not synchroinze the two (\textit{e.g.}\ we did not employ a boxcar integrator or lock-in amplifier of the measured voltage) an effect is required that lengthens our signal in time but does not accumulate between laser shots.
The qualitative features of our data, \textit{i.e.}\ 2PPV signal observed for 387 (or 774 nm) in combination with 3060 nm pulses but not 1530 nm pulses, are consistent with a scenario in which electron relaxation back to the Au between each laser pulse pair is exothermic and thus the local heat source. 
As shown in the Supporting information measurements with different electrolyte flow rates are consistent with a local heating signal transduction mechanism.   

\subsection{Vibration Assisted Charge Transfer}
As noted above, our data has three qualitative features (1) at negative delay times there is no signal (2) near time zero there is a spike with a timescale of $\approx$ 150 fs, and (3) a smaller signal persists for $\approx$ 4 ps after which there is no observable response. 
Prior work has clarified that femtosecond illumination of Au surfaces in contact with air initially creates an athermal population of high energy electrons that relaxes on three timescales: 10s of femtoseconds due to ballistic transport of hot electrons into bulk Au, a few hundred femtoseconds due to electron thermalization and a few picoseconds due to thermalization of the electronic and phononic subsystems \cite{hoh97,hoh00}. 

Our sample system differs from that of optically excited metal samples in air because there can be additional charge transfer pathways: \emph{hot} electrons may be transferred to the adsorbed FcC\textsubscript{11} monolayer or the aqueous electrolyte. 
Control experiments combining the 387 and 3060 nm pulses for an Au electrode without an FcC$_{11}$ SAM also find no observed 2PPV response (see our previous study \cite{Tong20}). 
Additionally no 2PPV response is observed for a system with 387 and 1530 nm excitation and an FcC$_{11}$ SAM. 
We conclude our signal results from charge transfer across the Au/electrolyte interface and requires \emph{\textbf{both}} that the FcC$_{11}$ SAM is present \emph{\textbf{and}} that its aromatic CH is vibrationally excited.  

As discussed above, the position of the open circuit voltage with respect to the Fc/Fc$^{+}$ current feature suggests the monolayer is Fc dominated in the absence of laser interaction. 
For all conditions in which a signal is observed, the 2PPV response is further negative (cathodic): the  monolayer is transiently further reduced. 
If the effect of the 387 or 774 nm pulses is to create high energy Au electrons in the language of \autoref{e:transition_states} they increase the over potential $\eta$.
However, given that electron transfer from Au to Fc$^{+}$ is already thermodynamically favored at open circuit voltages, transiently increasing $\eta$ is apparently insufficient to induce further electron transfer (recall no signal is observed when the 387 or 774 nm pulses precede the 3060 nm or at any delay in combination with 1530 nm pulses). 
We suggest the absence of charge transfer in these circumstances is the result of kinetics: solvent fluctuations are not sufficiently rapid to allow further electron transfer before the Au excitation has relaxed.
The 3060 nm excitation resonant with the aromatic CH  directly modifies $\lambda^{*}$, in the language of \autoref{e:transition_states}, but has a minimal effect on the energy of Au electrons.
Evidently, modulation of $\lambda^{*}$ is required to allow the transient increase in $\eta$ to produce an observable transient charge transfer (and thus further Fc$\rightarrow$Fc$^{+}$ reduction). 

Consistent with our signal resulting from laser modulation of the exponential in \autoref{e:transition_states}, see Supporting Information for data, in experiments that combine 387 and 3060 nm pulses the 2PPV signal is proportional to $\exp\left(I_{n}\right)$ where $I_n$ is the intensity of either beam. 
In experiments combining 774 nm and 3060 nm, the exponential dependence of the 2PPV response on mid infrared intensity is preserved but the 2PPV response is proportional to $\exp({(I_{774})^{1.7}})$ as expected if pulses of this photon energy interact with Au electrons via a two photon absorption.

\begin{figure}[htp]
	\begin{center}
		\includegraphics[width=0.5\textwidth]{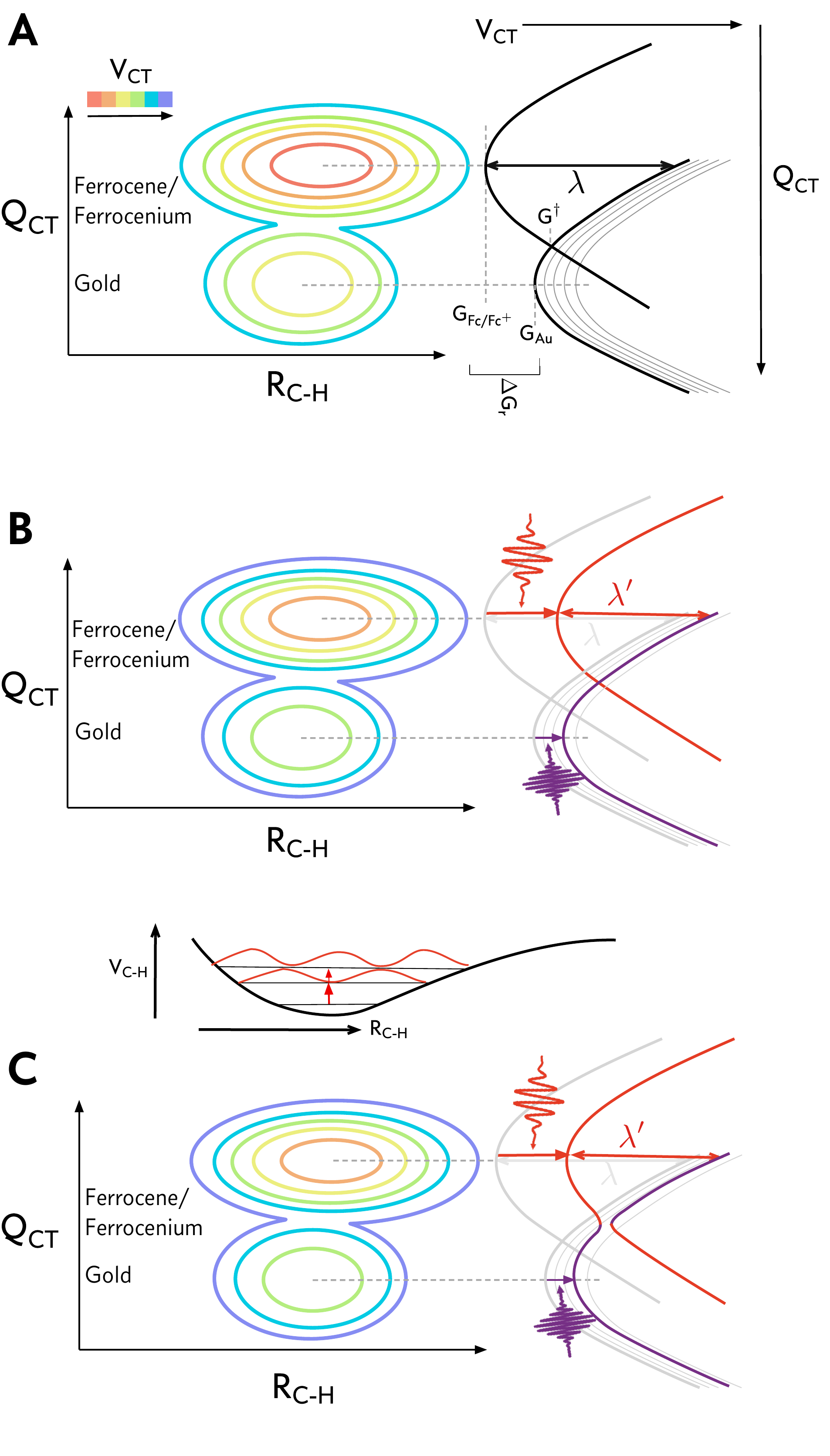}
	\caption{(A) The Au/FcC11 system at OCV before the arrival of the laser pulses represented in terms of the aromatic CH stretch bond distance (R$_{\text{C-H}}$) and the charge transfer coordinate (Q$_{CT}$). As suggested by the CV, electron transfer \emph{\textbf{to}} the Fc/Fc$^{+}$ state is favored thermodynamically. (B) As in (A) but now including both the visible (387 and 774 nm) and mid-IR (3060 nm) pulses. The former acts to increase the energy of Au electrons, the later to increase the energy of the FcC11 state (by decreasing $\lambda$). The charge transfer apparent in our experiment  at times > 500 fs is the result of kinetic factors, \textit{i.e.}\ $\lambda^{\prime} < \lambda$. (C) A scenario consistent with the large charge transfer signal we see on $\approx$ 150 fs timescales: coherent excitation of the aromatic CH stretch vibration acts to couple the Au donating and Fc/Fc$^{+}$ accepting states allowing charge transfer without the `vertical' transition in (A) and (B).    \label{fig:Scheme}}
	\end{center}
\end{figure}

These considerations rationalize our observation of no 2PPV signal at negative delays, 387 or 774 nm pulses before the 3060, and a 2PPV signal that has a 4 ps lifetime: the latter is set by the vibrational life time of the aromatic CH. 
They do not explain the large amplitude signal at timescales $\approx$ 150 fs.
In the weakly adiabatic regime electron donor and acceptor states are uncoupled. 
As a consequence (as shown in the right hand panel of \autoref{fig:Scheme}a), charge transfer occurs only `vertically' and requires overcoming $\lambda$.
Weak electronic coupling between the electron donor and acceptor states drastically increases electron transfer rates by creating an avoided crossing: an energy gap opens at the transition state and reactants can evolve towards products `horizontally' only overcoming G$^{\dag}$ (see right hand panel of \autoref{fig:Scheme}c).
Thus one way to rationalize the spike in our signal at short times is that it results from a short-lived increase in coupling between the donor Au states and the acceptor Ferrocenium.
Two physical effects may be responsible for such coupling: the initially athermal hot electron population may lead to enhanced charge transfer on $\approx$ 150 fs timescales or the in-phase aromatic CH stretch may result in a further enhancement of charge transfer in excess of that of a single vibrationally excited aromatic CH (vibrational dephasing is fast relative to vibrational relaxation).
It is worth emphasizing that if this short lived signal was the result of interfacial polarization we would expect to see such a response in measurements that combine either the 387 nm or 774 nm pulses and 1530 nm. 
The fact that we do not -- we only see a signal when combining visible pulses with 3060 nm -- suggests only charge transfer can rationalize our observations. 

The above discussion is summarized in \autoref{fig:Scheme}.
At equilibrium, \textit{i.e.} in the absence of laser fields, the position of the open circuit voltage cathodic of the oxidation current wave apparent in the CV suggests that reduction of Fc$^{+}$ to Fc, and thus Au$\rightarrow$Fc/Fc$^{+}$ electron transfer, is thermodynamically favored: $\Delta\text{G}_{\text{r}} = \text{G}_{\text{Au}} - \text{G}_{\text{Fc/Fc}^{\text{+}}} < 0$ as shown in \autoref{fig:Scheme}a. 
Both the visible (387 or 774 nm) pulse trains act to transiently \emph{increase} electron energy within the Au electrode thus \emph{increasing} the thermodynamic driving force for electron transfer from Au to Fc (as shown in \autoref{fig:Scheme}b). 
Yet our experiment combining either the 387 nm or 774 nm and 1530 nm pulses shows no 2PPV response.
This observation is consistent with a scenario in which electron transfer in such a system requires solvent fluctuations that are slow with respect to hot electron lifetime (\textit{i.e.}\ longer than ps). 
As shown in \autoref{fig:SFG} we expect experiments involving the 3060 nm pulse train to resonantly excite aromatic CH stretch vibrations on the Ferrocene moiety. 
Because such light/matter interaction creates a non-equilibrium population with respect to the CH stretch potential we expect it to \emph{also} result in an increase in energy of the FC/FC$^{+}$ state:   $\Delta G_{r}$ of Au$\rightarrow$Fc/Fc$^{+}$ charge transfer becomes less negative.     
The fact that we observe a charge transfer response for experiments containing either 387 or 774 nm pulse trains \emph{and} the 3060 nm suggests that under these conditions the reaction Au$\rightarrow$Fc$^{+}$ electron transfer is still favorable, \textit{i.e.}\ $\Delta \text{G}_{\text{r}} < 0$, but that the decrease in solvation energy, \textit{i.e.}\ $\lambda^{\prime} < \lambda$, is sufficient to allow it to proceed within the hot electron lifetime. 
This scenario is illustrated in \autoref{fig:Scheme}B.
\autoref{fig:Scheme}C illustrates a scenario in which coherent excitation of the aromatic CH stretch vibration correlates with weak coupling between donor and acceptor states along the charge transfer coordinate.

Taken together, it seems clear that vibrational excitation of the aromatic CH stretch acts to significantly enhance charge transfer over timescales similar to its vibrational lifetime. 
In this context it is worth noting that the energy of this aromatic CH vibration, 0.38 eV, is a significant fraction of the $\lambda_{\text{eff}}$ for the bicomponent FcC11 SAM estimated from electrical perturbation measurements by Chidsey \cite{chi91}: 0.85 eV.
This correspondence is consistent with a scenario in which $\lambda_{\text{eff}}$ inferred from electrochemical measurements results from a small number of solvent fluctuations ($\nu$ in \autoref{e:lambda} is small).
Clearly under such conditions the approximation of reactants or products interacting with solvent through single harmonic oscillator of energy $\lambda_{\text{eff}}$ bears little resemblance to the underlying microscopic physics.

A number of workers have recognized that a \emph{single} effective $\lambda$ is, for the vast majority of reactions, a profound oversimplification. 
Many extensions have been proposed.
For example, originating in early work from Marcus \cite{mar56}, a $\lambda_{fast}$ capturing reactant/product polarization and a $\lambda_{slow}$ describing nuclear rearrangement.
Along similar lines it has been recognized that reactants and product need not have the same $\lambda$s (\textit{e.g.}\ see the discussion of Schmickler and Santos \cite{sch10}).  
More recently Huang has developed a theoretical approach describing the electronic effect of the surface on $\lambda$ \cite{hua20b} and several groups have addressed the possibility of solvents that interact via nonlinear potentials with reactants or products \cite{mat09,mat17,lu20}. 
Such treatments require relaxation of the assumed \emph{ergodic} solvent bath, \textit{i.e.}\ they allow the possibility that some subsection of solvent fluctuations may be slow relative to charge transfer. 
However such approaches do not allow the possibility of reactant, transition state or product coupling to \emph{\textbf{particular}} modes in the electrolyte (in the language of \cite{lu20} and \autoref{e:lambda} the reactant/solvent coupling is not frequency dependent).
As a result such theories do not allow a convenient way to address trends in reactivity with electrolyte (they elide specific interactions between electrolyte and reactant or product) and are inconsistent with our results.

\section{Summary and Conclusions}
In this study we demonstrate vibrationally assisted charge transfer across a prototypical electrode/aqueous electrolyte interface: FcC$_{11}$ self-assembled monolayers on Au in contact with an aqueous electrolyte.
We do so by showing that in the presence of optical pulses exciting the underlying Au electrode resonant excitation of the aromatic CH stretch in the FcC$_{11}$ SAM induces charge transfer on femtosecond timescales while non-resonant laser interactions do not. 
This measurement is the only type of which we are aware that can offer insight into the right hand side of \autoref{e:lambda}: the only way to experimentally elucidate the manner in which \emph{\textbf{specific}} solvent motions interact with charge transfer.

(Semi-)analytic theories of heterogeneous charge transfer, in systems with chemically interesting adsorbate/electrode interaction, typically suggest that the reactant (or product) has a single effective interaction with solvent (in which this interaction may be nonlinear). 
Such an effective interaction is consistent with a scenario in which a limited number of distinct solvent motions contribute to the charge transfer coordinate. 
The energy of the aromatic CH stretch, 0.38 eV is a significant fraction of the $\lambda_{\text{eff}}$ inferred from electrochemical measurements of similar systems: 0.85 eV. 
Thus our results suggest that evolution of this system along its charge transfer coordinate may, in fact, be dominated by a relatively small number of solvent structural fluctuations. 
 
Clear application of this approach to other electrochemical systems -- particularly systems where electrolyte composition is known to dramatically effect reactivity -- are also necessary. 
Nevertheless this study offers a road map for the new insights into solvent control on heterogeneous transfer. 
While the identification of \emph{particular} solvent fluctuations that contribute to charge transfer seems likely to be system specific, our work suggests it would be productive to explicitly investigate the consequences of reactants or products interacting with limited numbers of solvent fluctuations, widely separated in frequency, on predicted charge transfer rates and offers an experimental approach to the exploration of these relationships.
Such insights are required to optimize electrocatalytic activity via \emph{electrolyte engineering}.

\section{Acknowledgements}
    This work was supported by Deutsche Forschungsgemeinschaft (DFG, German Research Foundation) through Germany's Excellence Strategy EXC 2033-390677874-RESOLV and through Project No. 278162697 - SFB 1242.
    Support by the European Research Council under the European Union's Horizon 2020 research and innovation program (grant agreement no. 772286, to RKC) is also gratefully acknowledged.

\section{Supporting Information}
Please see the Supporting Information for characterization of the output of the OPCPA system; description of the VSFG spectrometer; cyclic voltammetry results; a detailed discussion of the processing of the 2PPV data and a demonstration that our results are insensitive to the details of the method chosen; a description of a variety of additional control experiments; measurements illustrating the fluence dependence of the 2PPV response; and two beam cross-correlations for the 387 + 3060 nm and 774 + 3060 nm experiments.

\bibliographystyle{ieeetr}
\bibliography{Ferrocene_Bibliography}

\end{document}